\documentclass[prl,twocolumn,floatfix,showpacs,superscriptaddress]{revtex4}
\usepackage{graphicx}
\usepackage{amsmath}
\begin{document}

\title{Topology induced Kondo effect in hydrogenated Pt nanocontacts}
\author{G. Chiappe}
\affiliation{Departamento de F\'{\i}sica Aplicada, Unidad Asociada del
Consejo Superior de Investigaciones Cient\'{\i}ficas and
Instituto Universitario de Materiales, Universidad
de Alicante, San Vicente del Raspeig, Alicante 03690, Spain.}
\affiliation{Departamento de F\'{\i}sica J.J. Giambiagi, Facultad de
Ciencias
Exactas, Universidad de Buenos Aires, Ciudad Universitaria,
1428 Buenos Aires, Argentina.}
\author{E. Louis}
\affiliation{Departamento de F\'{\i}sica Aplicada, Unidad Asociada del
Consejo Superior de Investigaciones Cient\'{\i}ficas and
Instituto Universitario de Materiales, Universidad
de Alicante, San Vicente del Raspeig, Alicante 03690, Spain.}
\author{E.V. Anda}
\affiliation{Departamento de F\'{\i}sica, Pontificia Universidade
Cat\'olica
do R\'{\i}o de Janeiro, 22452-970 R\'{\i}o de Janeiro, Caixa Postal 38071,
Brasil.}
\author{J. A. Verges}
\affiliation{Departamento de Teor\'{\i}a de la Materia Condensada, Instituto de Ciencias de Materiales de Madrid (CSIC), Cantoblanco, Madrid 28049, Spain}
\date{\today}
\begin{abstract}
It is shown that recent experimental data on electronic transport through
Pt nanocontacts in the presence of hydrogen admit an explanation
in terms of topological and electron-electron correlation grounds.
A model Hamiltonian, which incorporates two orbitals on Pt atoms,
a single orbital on hydrogens, and the on-site Coulomb
repulsion  on the H atoms, is solved exactly, and connected to Pt leads
described  by a Bethe lattice.  When two weakly coupled H atoms
are placed between the Pt electrodes transversally
to the transport direction (as recently suggested) a  Kondo effect
related to the symmetry of the Pt-H couplings, stabilizes the conductance
around one quantum with a single channel contributing to the current,
in agreement with the experiments.
\end{abstract}
\maketitle

\begin{figure}
\includegraphics[width=2.5in,height=1.4in]{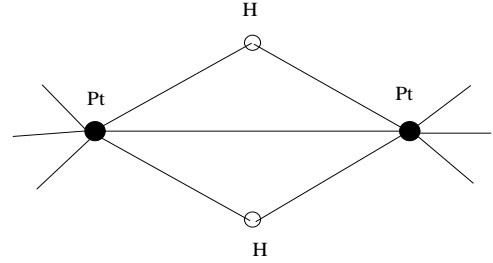}
\caption{
Sketch of the geometry of the Pt$_2$H$_2$ cluster connected to bulk Pt
(described by means of a Bethe lattice, see text).  
The $z$ axis is taken perpendicular to the cluster plane.  
\label{fig:structure}}
\end{figure}

A recent experimental work  reports that electronic
transport  through Pt nanocontacts is strongly modified in the
presence of hydrogen \cite{SN02,Sm03}. Specifically, while in
Pt nanocontacts the main conductance peak lies close to 1.5
conductance quanta ${\mathcal G}_0=2e^2/h$, in a hydrogen atmosphere the
main peak shows up near one quantum. Although
it was suggested \cite{SN02,Sm03} that the latter result was due to a
H molecule being placed along the nanocontact axis bridging
the last two Pt atoms of the left and right electrodes, an
alternative explanation has been put forward recently \cite{GP04}.
Extensive quantum chemistry calculations led  the authors
to conclude that a geometry in which two
weakly coupled hydrogen atoms were placed transversally
to the nanocontact axis (see Fig.\ref{fig:structure}),
was  more probable.  Most of the results described in \cite{SN02}
found a reasonable explanation  \cite{GP04}, including vibrational
frequencies and the isotope effect, but the observed unit conductance
to which a single channel apparently contributes.
The aim of this work is to study the strong
correlation effects that one may expect due to the presence of hydrogen, 
the role played by topology, recently emphasized in an analysis 
\cite{AD04} of transport through two quantum dots in a similar geometry (dots
located transversally to the nanostructure axis), and the importance of both in
understanding the experimental results \cite{SN02}.

The model Hamiltonian used to describe the Pt-H-H-Pt cluster shown
in Fig. \ref{fig:structure} is:
\begin{equation}
{\hat H} = {\hat H_{Pt}}+{\hat H_H}+{\hat H_{Pt-H}}
\end{equation}
\noindent The Pt Hamiltonian includes two isotropic ($s$-like) orbitals,
the first representing the
6s Pt orbital, and the second some average of the five 5d Pt orbitals,
\begin{eqnarray}
{\hat H_{Pt}} & = &\sum_{i=1,2;\sigma}\left[\epsilon_a
a^{\dagger}_{i\sigma}
a_{i\sigma}+\epsilon_b b^{\dagger}_{i\sigma}b_{i\sigma}
\right] +\nonumber  \\
&&\sum_{<ij>;\sigma}\left[t_{aa}a^{\dagger}_{i\sigma}
a^{\dagger}_{j\sigma}+
t_{bb}b^{\dagger}_{i\sigma}b_{j\sigma} \right]
\label{eq:H_Pt}
\end{eqnarray}
\noindent As in the geometry of Fig.\ref{fig:structure} the H atoms
are rather far appart  \cite{GP04} we do not include the
interaction between their respective s-orbitals. On the other hand, the
Hamiltonian incorporates the on-site Coulomb repulsion through a Hubbard
term,
\begin{equation}
{\hat H_H}  = \sum_{i=1,2;\sigma}
\epsilon_h h^{\dagger}_{i\sigma} h_{i\sigma} +
U_hh^{\dagger}_{i\uparrow}h_{i\uparrow}
h^{\dagger}_{i\downarrow}h_{i\downarrow}
\label{eq:H_H}
\end{equation}
\noindent Finally the Pt-H interactions are written as,
\begin{equation}
{\hat H_{Pt-H}}  = \sum_{i,j=1,2}(t_{ha}h^{\dagger}_{i\sigma} a_{j\sigma}+
t_{hb}h^{\dagger}_{i\sigma} b_{j\sigma})
\label{eq:H_Pt-H}
\end{equation}
\begin{table}
\caption{
Energies of the  Pt and H orbitals and the interactions amongst them
(all in eV), as defined in Eqs. (1-4), used to describe the  Pt$_2$ and the
Pt$_2$H$_2$ clusters. Parameters for bulk Pt were
taken equal to those of the Pt$_2$ cluster.}
\label{Table}
\begin{tabular}{|c|c|c|c|c|c|c|c|c|}
\colrule
  & $\epsilon_a$ & $\epsilon_b$  & $\epsilon_h$ & $t_{aa}$ & $t_{bb}$ &
  $t_{h_{i}a_{j}}$  & $t_{h_{i}b_{j}}$  & $U_h$\\
 \colrule
 bulk Pt & -7.0 & -7.5 & - & -1 & 0.2 & - & -  & -\\  
  \colrule
 Pt$_2$H$_2$ & -6.75 & -7.75 & -9.5 & -1.2 & 0.2 & -0.5 & (-1)$^{i+j}$ 0.5  & 7.0 \\
\colrule
\end{tabular}
\end{table}
When the cluster is connected to electrodes, the transmission across the system is given by \cite{La57}:
\begin{equation}
T(E)=\frac{2e^2}{h}{\rm Tr}[t^{\dagger}t].
\label{g}
\end{equation}
\noindent and the conductance is ${\mathcal G}=T(E_F)$, where $E_F$ is
the Fermi level that will be hereafter taken as the zero of energy.
In this expression,  Tr denotes the trace over all  orbitals in the cluster
and the matrix $t$ is given by
\begin{equation}
t= \Gamma_{\rm R}^{1/2}G^{(+)}\Gamma_{\rm L}^{1/2}=
\left[\Gamma_{\rm L}^{1/2}G^{(-)} \Gamma_{\rm R}^{1/2} \right]^{\dagger},
\end{equation}
where the  matrix $\Gamma_{\rm R(L)}$
is given by $i(\Sigma^{(-)}_{\rm R(L)}-\Sigma^{(+)}_{\rm R(L)})$,
$\Sigma^{(\pm)}_{\rm R(L)}$  being the  
self-energies of the right (R) and left (L) leads.
The Green function  is written as,
\begin{equation}
G^{(\pm)}=\left(\left[G_0^{(\pm)}\right]^{-1}-c
\left[\Sigma_R^{(\pm)}+\Sigma_L^{(\pm)}\right]\right)^{-1}
\label{eq:green}
\end{equation}
where $G_0^{(+)}$ is the Green function
associated to the isolated Pt$_2$H$_2$ cluster, which is obtained
by means of exact diagonalization using the Lanczos method \cite{FC99,CV04}.
The bulk of the electrodes is described by means of a Bethe lattice with coordination 
twelve, while the two Pt in the cluster are connected to $c$ bulk atoms. 
As discussed below, this effective coordination $c$ is taken as a fitting parameter. 
In order to single out the contribution of individual channels we
diagonalize the matrix $t^{\dagger}t$.
This method, hereafter denoted as the Embedded Cluster Method (ECM), is {\it exact} 
only as far as the calculation of the Green function of the isolated cluster is 
concerned, and does not account for electron-electron interaction effects that 
may extend beyond its bounds.
In the Kondo regime, many-body effects extend over a
length given by the Kondo cloud that is proportional to the inverse of
the atomic Kondo temperature $T_k$.
In this case, as hydrogen and Pt interact strongly, we expect this
temperature to be large enough to guarantee a rather localized  cloud. 
Aiming to highlight the role played by the Kondo effect we also solve this
system using Unrestricted Hartree Fock (UHF) that gives an appropriate
description for temperatures greater than the Kondo temperature.
\begin{figure}
\includegraphics[width=2.5in,height=3.2in]{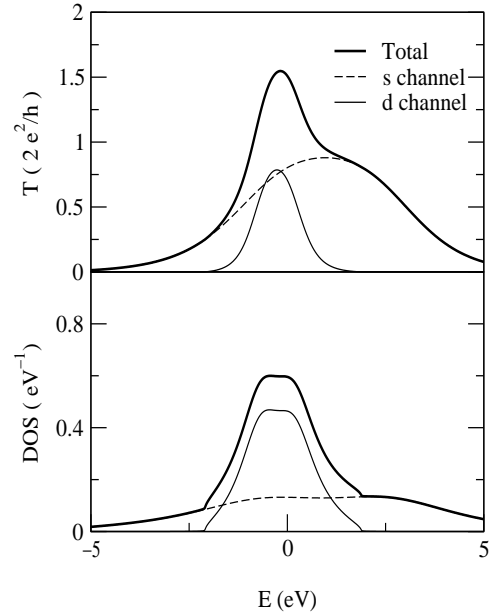}
\caption{
Upper panel: transmission as a function of  energy for the $Pt_2$ cluster.
Total transmission (thick continuous line) and the contributions of the
$d$ i(continuous line) and $s$ (broken line) eigenchannels are shown.
Lower panel: same for the density of states (DOS).  
\label{fig:G_Pt}}
\end{figure}

The choice of the model parameters (given in Table I) was guided by
{\it ab-initio} calculations of the Pt$_2$ and Pt$_2$H$_2$ clusters.
As that calculation indicates that the $d_{xy}$ orbital is the one that
more strongly interacts with the H orbital, orbital $b$ in  
Eq. (2) was assumed to have that symmetry in the Pt$_2$H$_2$ cluster.
The parameters of Table I give a bulk Bethe-lattice Density of States (DOS)
with a $d$ band  more than half-filled and  a $s$ band less than
half-filled.
All Bethe-lattice and nanocontact calculations were carried out by
fixing the charge to one (two) electron per H (Pt) atom.  
Parameter $c$ in Eq. (\ref{eq:green}) was chosen to
give the experimental conductance in Pt nanocontacts with no hydrogen
\cite{SN02,GP04}.  

\begin{figure}
\includegraphics[width=2.5in,height=3.2in]{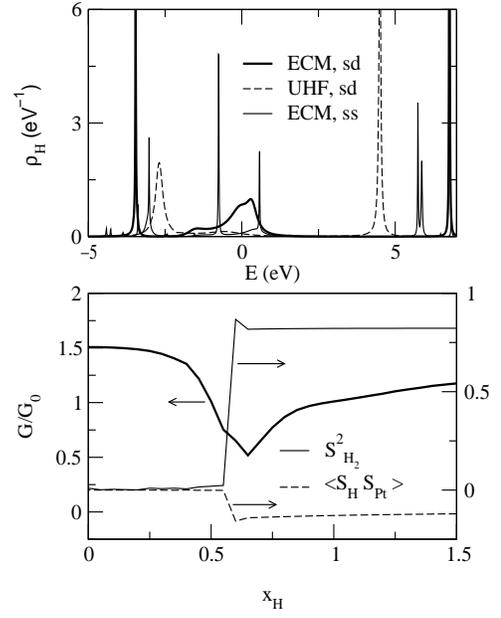}
\caption{Upper panel: Local DOS at the H in the $Pt_{2}H_{2}$ nanocontact,
as calculated by means of  UHF and ECM,  with either two
$s$ Pt orbitals or two orbitals of $s$ and $d_{xy}$ symmetry (see text).
Lower panel: Conductance and spin correlations calculated with ECM as
function of the degree of hydrogenization (see text).
\label{fig:DOSH}}
\end{figure}

Fig.\ref{fig:G_Pt}
shows our results for the transmission and the DOS in this
nanocontact (with $c$=5.5), which
are in line with those reported in \cite{GP04}. {\it Both channels equally contribute} 
to give a conductance around ${1.6\mathcal G}_0$. Due to the symmetry of the $d_{xy}$ 
orbital there is no indirect (hydrogen mediated) $s-d$ coupling between $Pt$ atoms.  
In addition, direct coupling between $s$ and $d$ orbitals on different
atoms is small enough  to be ignored in our simplified model.
As a result, $s$ and $d$ orbitals act independently even in the presence
of hydrogen.
\begin{figure}
\includegraphics[width=3.4in,height=3.2in]{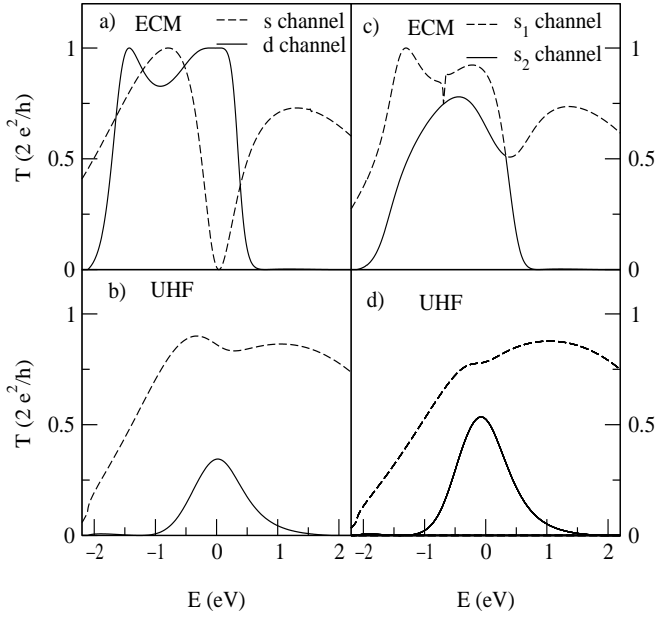}
\caption{Left panels: $s$ and $d$ transmissions versus energy in the
Pt$_2$H$_2$ cluster as calculated with ECM (upper) and UHF (lower).
Right panels: same but replacing the Pt $d$ orbital by an s-like
isotropic orbital.
\label{fig:G_PtH}}
\end{figure}
Our results for the transmission in the hydrogenated Pt nanocontact are
depicted in Fig.\ref{fig:G_PtH}. Comparing these results with those
without H (see Fig.\ref{fig:G_Pt}) we note that now
the transmission reaches a maximum (minimum) of around  ${\mathcal
G}_0$ (0) for the $d$ ($s$) eigenchannel. This result is not reproduced by the UHF 
approximation, as shown in  the lower panel of Fig.\ref{fig:G_PtH}.  
The latter resembles the transmission for the Pt nanocontact
(see Fig.\ref{fig:G_Pt}), but reduced by the effect
of scattering with $H$ orbitals. 
The local DOS at the H atoms is shown in the upper panel of
Fig.\ref{fig:DOSH}. The ECM result  shows  
the Kondo signature at the Fermi level, which is absent in the UHF
calculation, that describes the solution for $T > T_k$.
The comparison of UHF and ECM results clearly suggest the
presence of strong many-body effects. As there is no direct or indirect
hybrization between $s$ and $d$ bands, the pair  of H can
be in the Kondo regime with each Pt band simultaneously and independently,
as described in \cite{AD04} for a one band model. In this regime, both
hydrogens become strongly correlated (in a state of total spin
$S_{H_2}=1$), forming a Correlated Impurity of Spin 1 (CIS1), side connected 
to the respective channel in our model. Then, the conductance in each
channel has two contributions coming from: i) direct coupling between
$d$ or $s$ orbitals, and, ii) coupling mediated by the CIS1.
Charge in the the CIS1 induce a phase shift of $\pi$ respect to the
direct contribution, as can be argued from the Friedel sum rule
\cite{note1}, and was also verified numerically by us. Contributions from both 
ways are 
similar and  therefore a destructive interference take place in the s channel. 
In the case of the $d$ channel there is an additional $\pi/3$
phase shift, because the H are connected with different sign to each  
Pt. This prevents destructive interference in this channel.
Moreover, in this case the dominant contribution to the conductance
comes from the impurity channel (see Table I) that is in the Kondo regime 
given conductance very close to $1$.
 Results for the transmission and local DOS for the model
with two $s$-like orbitals ($s_1$ and $s_2$) in place of $s$ and $d$
orbitals are also shown in Figs.\ref{fig:G_PtH} and \ref{fig:DOSH}.
In this case there is no Kondo resonance at the Fermi level, as
$s_1$ and $s_2$ orbitals on different Pt atoms are now coupled through the
hydrogens. As shown in Fig.\ref{fig:DOSH}, this coupling splits the Kondo
resonance into two peaks above and below the Fermi level.
As a consequence,  only the direct coupling between $Pt$ atoms
contribute to the conductance and, as interference effects
do not show up, the two channels give
similar contributions to the conductance (see Fig.\ref{fig:G_PtH}).
This  is also the case
of a $H_2$ molecule placed longitudinally to the nanocontact axis, as,
in this case, each $H$ atom is connected only either to the left or  right
electrodes, giving an indirect coupling between $s$ and $d$ bands.  

In order to investigate how the conductance changes from a situation
with no hydrogen up to that where the two H atoms are located as in
Fig.\ref{fig:structure}, we have carried out calculations
varying the H-Pt parameters as $t_{hi}^{'}= x_H t_{hi}$, where $i=a,b$ 
and $x_H$  varying from $0$ (H free nanocontacts) up to $1.5$.
The most appealing feature of the results shown in the lower panel of 
Fig.\ref{fig:DOSH}  is the wide range of $x_H$ over which the ECM  gives a 
conductance close to one \cite{note2}.
This is due to  the
electron-electron interactions that are playing an essential
role in this system. The latter statement is corroborated
by the spin-spin correlations also shown in Fig.\ref{fig:DOSH}.
When the hydrogen atoms are not coupled (or just weakly coupled) to Pt,
H-H correlations are antiferromagnetic and the total spin of the
molecule is zero, as occurs in the free hydrogen molecule. In this limit Pt-H
spin-spin correlation is zero, as it should. However, just when the
conductance stabilizes around one, H-H spin correlations  
become strongly ferromagnetic, and the total spin of the molecule
raises up to $0.8$, while the Pt-H bond shows antiferromagnetic
correlations. These results  illustrate the many features that
this many-body effect shares with the standard Kondo effect
of spin 1 already obtained in a two parallel quantum dot system
\cite{AD04}.
\begin{figure}
\includegraphics[width=2.5in,height=3.2in]{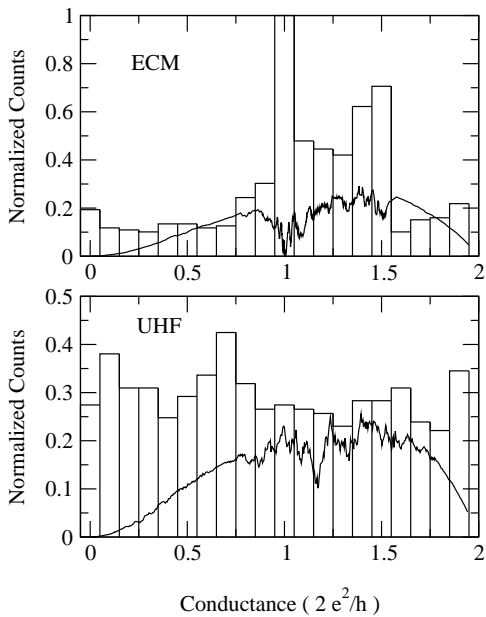}
\caption{
Normalized conductance histogram and normalized conductance fluctuations
for the hydrogenated nanocontact versus the conductance through it.
The results were obtained by using either ECM (upper
panel) or UHF (lower panel).
\label{fig:fluctuations}}
\end{figure}

The method used in \cite{SN02} to estimate the number of channels that
contribute to the current consists of calculating the conductance
fluctuations
obtained in a large number of  experiments. The experimental results
indicate that the derivative of the conductance $G$ with respect to the
bias voltage $\partial G/\partial V$ varied randomly obeying
a bell-shaped distribution whose standard deviation was calculated in
\cite{LD99},
\begin{equation}
\sigma_{GV} \propto \sqrt{\sum_{n=1,..,N} T_n^2(1-T_n)}
\label{eq:model}
\end{equation}
\noindent where  $T_n$ is the transmission probability of channel $n$ and
$N$ is the total number of channels.
We have calculated  the conductance fluctuations by randomly varying
$x_H$ over the range [0,1.5]. Couplings between $Pt$ atoms in the nanocontact
were also multiplied by a random factor in the range 0-1.5.
Averages were done over at least 1000 realizations.
To calculate the standard deviation, instead of directly working on the
distribution $\partial G/\partial V$ that gives very noisy results, we
introduced the  results for the transmission of each channel into  
Eq. (\ref{eq:model}).  As shown in \cite{LP04}  both
calculations agree in identifying the minima in fluctuations.
The ECM results depicted in
Fig.\ref{fig:fluctuations} clearly show a  peak in the
conductance histogram near  ${\mathcal G}_0$,
with a minimum in fluctuations. This  means that only one channel
is contributing to the unit conductance, being this value  
the most probable.
The UHF results, however, show finite fluctuations  over the whole range.

Summarizing, the effects of electron-electron interaction on transport
through hydogenated Pt nanocontacts have
been investigated. Our results indicate that when two H are located 
transversally to the nanocontact axis \cite{GP04}, a Kondo effect stabilizes 
the conductance around one quantum. 
Spin-spin correlations are much like those which occur in the standard Kondo 
effect observed on two dots in a similar geometry \cite{AD04}.  The effect 
occurs due to the decoupling of $d$ and $s$ bands produced by the
combination of the transversal location of the molecule and to the $d_{xy}$
symmetry of the Pt d-orbital.
The result is robust and largely independent of the model parameters.
Besides, a single channel contributes to the conductance, leading to
a minimum in the conductance fluctuations, in agreement with
experiments \cite{SN02}. This analysis permits
to conclude  that the unit conductance observed in this system is the
result of a combination of highly Kondo correlated spins and topology.

\acknowledgments
Financial support by the spanish MCYT (grant MAT2002-04429-C03 and
a sabatical grant), the Universidad de Alicante, the Universidad de Buenos 
Aires (grant UBACYT x210) and Conicet, is gratefully acknowledged.
Useful  discussions with J. Fern\'andez-Rossier, Y. Garc{\'\i}a, J.J. Palacios,
A. P\'erez-Jim\'enez, and E. San Fabi\'an are  gratefully acknowledged.


\begin{thebibliography}{99}
\bibitem{note1}
The Friedel sum rule can be used to calculate the phase shift in
each spin channel accross an interacting system. It gives $\pi n_\sigma$,
where $n_\sigma$ is the charge per spin at the impurity. In our case
$n_\sigma=1/2$ on each hydrogen, which gives $\pi$ for the phase shift
(see {\protect \cite{He93}}).
\bibitem{note2}
The Kondo-like peak that shows up in quantum dots only gives unit
conductance at half-filling and in the presence of electron-hole
symmetry (see, for instance, {\protect \cite{AD04}}).
This is not the case in the present system.
\end{thebibliography}
\end{document}